\documentclass[manuscript]{aastex}
\usepackage{graphicx}
\usepackage{txfonts}
\usepackage{graphicx}
\usepackage{mathrsfs}

\slugcomment{Accepted to ApJL}

\shorttitle{
	Gaseous CO Abundance---An Evolutionary Tracer for Molecular Clouds} \shortauthors{Liu et al.}

\begin{document}

\title{Gaseous CO Abundance---An Evolutionary Tracer for Molecular Clouds}
\author{Tie Liu\altaffilmark{1}, Yuefang Wu\altaffilmark{1}, Huawei Zhang\altaffilmark{1}}

\altaffiltext{1}{Department of Astronomy, Peking University, 100871,
Beijing China; liutiepku@gmail.com, ywu@pku.edu.cn }

\begin{abstract}
Planck cold clumps are among the most promising objects to investigate the initial conditions of the evolution of molecular clouds.
In this work, by combing the dust emission data from the survey of Planck satellite with the molecular data of $^{12}$CO/$^{13}$CO (1-0) lines from
observations with the Purple Mountain Observatory (PMO) 13.7 m telescope, we investigate the CO abundance, CO depletion and CO-to-H$_{2}$ conversion factor
of 674 clumps in the early cold cores (ECC) sample. The median and mean values of the CO abundance are 0.89$\times10^{-4}$ and 1.28$\times10^{-4}$, respectively.
The mean and median of CO depletion factor are 1.7 and 0.9, respectively. The median value of $X_{CO-to-H_{2}}$ for the whole sample is $2.8\times10^{20}$ cm$^{-2}$K$^{-1}$km$^{-1}$~s. The CO abundance, CO depletion factor and CO-to-H$_{2}$ conversion factor are strongly (anti-)correlated to other physical parameters (e.g. dust temperature, dust emissivity spectral index, column density, volume density and luminosity-to-mass ratio). To conclude, the gaseous CO abundance can be used as an evolutionary tracer for molecular clouds.
\end{abstract}


\section{Introduction}
Carbon monoxide (CO) is the second most abundant molecular
species (after H$_{2}$) in molecular clouds and is often used to determine the column density of molecular hydrogen by assuming a [CO/H$_{2}$] abundance ratio. Previously different authors used different [CO/H$_{2}$] abundance ratio \citep{wu04}, which is from 2.5$\times10^{-5}$ \citep{rod82} to 10$^{-4}$ \citep{gar91}. In addition, the observations of $^{12}$CO (1-0) are often used to estimate the molecular content of entire galaxies by applying an empirical CO-to-H$_{2}$ conversion factor ($X_{CO-to-H_{2}}=N_{H_{2}}/I_{CO}$), which is (1.8$\pm0.3)\times10^{20}$ cm$^{-2}$K$^{-1}$km$^{-1}$~s for the disk of the Milky Way \citep{dame01}. However, the CO-to-H$_{2}$ conversion factor varies with different methods used in measuring the column density of H$_{2}$ \citep{pine08}. The reliability of using CO as a tracer for molecular mass should be taken seriously because the abundance of gaseous CO is very sensitive to chemical effects such as CO depletion in cold regions.

Towards five low-mass starless cores, \cite{taf02} found that the abundance of CO near the core center decreases by at least 1 or 2 orders of magnitude with respect to the value in the outer core, indicating that huge amount of gaseous CO freezes out onto dust grains in the dense region. In the observations towards 21 IRDCs, the CO depletion factor ($f_{D}$), which is defined as the ratio of expected standard gas phase CO abundance to the observed CO abundance, is in between 5 and 78, with a median value of 29 \citep{fon12}. Additionally, the depletion of gas-phase CO seems to increase with density. \cite{case99} found that the depletion factor is up to $\sim10$ where the mass surface density is $\Sigma\simeq0.6$ g~cm$^{-2}$, while \cite{kram99} found the depletion factor is $\sim2.5$ for regions with $\Sigma\simeq0.1-0.15$ g~cm$^{-2}$. In the observations towards the filamentary IRDC
G035.30-00.33, \cite{her11} found the depletion factor increases by about a factor of five as $\Sigma$ increases from $\sim$0.02 to $\sim$0.2 g~cm$^{-2}$. By combining data from the Five College Radio Astronomy Observatory CO Mapping Survey of the Taurus molecular cloud with extinction data for a sample of 292 background field stars,
\cite{whi10} found that the mean ratio of icy CO to gaseous CO increases monotonically from negligible levels for visual extinctions $A_{V}\leq5$ to $\sim0.3$ at $A_{V}=10$ and $\sim0.6$ at $A_{V}=30$. However, in the survey towards the Gould Belt clouds, only in the cases of starless cores in Taurus and protostellar cores in Serpens, there is a correlation between the column densities of the cores and the depletion factor \citep{chr12}. And similarly, CO depletion factor does not seem to be correlated to any other physical parameter in the observations towards 21 IRDCs \citep{fon12}.

However, previous works are severely subject to relatively small sample and thus they can not statistically investigate the relationships between gaseous CO abundance and the other physical parameters. In this paper, we use the early cold cores (ECC) sample to systemically investigate the gaseous CO abundance, CO depletion and CO-to-H$_{2}$ conversion factor in molecular clumps. Our results suggest that the gaseous CO abundance strongly (anti-)correlates with dust temperature, dust emissivity spectral index, column density, volume density and luminosity-to-mass ratio.

\section{Data}
The early cold cores (ECC) sample is a subset of the Planck Early Release Compact Source Catalogue and contains only the most reliable detections (SNR$>$15) of sources with color temperature below 14 K \citep{ade11}. In the ECC, the photometry is carried out on the original Planck maps by
placing an aperture of 5$\arcmin$ radius on top of the detection \citep{ade11}. The background
is estimated using an annulus around the aperture with an
inner radius of 5$\arcmin$ and an outer radius of 10$\arcmin$. Temperatures and dust emissivity spectral indexes are
derived from a fit to all four bands (IRIS 3 THz and Planck 857,
545 and 353 GHz) \citep{ade11}. The major and minor axis of each source are also obtained from ellipse fit \citep{ade11}. We extracted the aperture flux density at 857 GHz, the core temperature, core emissivity index as well as the ellipse major and minor axis of each clump from the ECC catalogue.

We have carried out a follow-up observations towards 674 ECC clumps with the Purple Mountain Observatory (PMO) 13.7 m telescope. The details of the observations can be found in \cite{wu12}. The half-power beam width of the telescope at the 115 GHz frequency
band is 56$\arcsec$. The main beam efficiency is $\sim$50\%. For each identified $^{13}$CO (1-0) component, their kinematic distance and galactocentric distance were calculated \citep{wu12}.

\section{Results}
\subsection{Abundance of gaseous $^{12}$CO}
The excitation temperature of $^{12}$CO (1-0) can be derived as following \citep{gar91}:
\setcounter{equation}{0}
\begin{equation}
T_{r}=\frac{T_{a}^{*}}{\eta_{b}}=\frac{h\nu}{k}\left[\frac{1}{exp(h\nu/kT_{ex})-1}-\frac{1}{exp(h\nu/kT_{bg})-1}\right]\times\left[1-exp(-\tau)\right]f
\end{equation}
here $T_{a}^{*}$ is the antenna temperature, and $T_{r}$ is the brightness temperature corrected with beam
efficiency $\eta_{b}$, h is the Planck constant, k is the Boltzmann constant
and $\nu$ is the frequency of observed transition. Assuming $^{12}$CO (1-0) emission is optically thick ($\tau\gg1$) and the filling factor f=1, the excitation temperature $T_{ex}$ can be
straightforwardly obtained. Since the Planck cold clumps are clearly extended and are mostly resolved by Planck observations \citep{ade11}, the assumption of filling factors for gas f=1 is reasonable.

Assuming the $^{13}$CO (1-0) and C$^{18}$O (1-0) lines are optically thin and they have the same excitation temperature as $^{12}$CO (1-0), the column density of $^{13}$CO and C$^{18}$O under LTE condition can be obtained as following \citep{gar91,pill07}:
\begin{equation}
N_{total}=\frac{3h\varepsilon_{0}}{2\pi^{2}S\mu_{g}^{2}}\frac{J(T_{ex})Q(T_{ex})W}{J_{\nu}(T_{ex})-J_{\nu}(T_{bg})}
\end{equation}
where $\varepsilon_{0}$ is the dielectric permittivity, $S\mu_{g}^{2}$
is the line strength multiplied by the dipole moment along the
molecular g-axis. $J(T_{ex})$ is defined as \citep{pill07}
\begin{equation}
J(T_{ex})=\frac{exp(E_{u}/kT_{ex})}{exp(h\nu/kT_{ex})-1}
\end{equation}
where $E_{u}$ is the upper energy level. $J_{\nu}(T)$ is defined as \citep{pill07}
\begin{equation}
J(T)=\frac{h\nu/k}{exp(h\nu/kT)-1}
\end{equation}
where $T_{bg}=2.73$ is the temperature of the cosmic background
radiation.

Then we calculated the peak optical depth $\tau_{0}$ of $^{13}$CO (1-0) and C$^{18}$O (1-0) from equation (1) and applied a correction factor $C_{\tau-LTE}=\frac{\tau_{0}}{1-exp(-\tau_{0})}$ to the column density. The correction factor obtained using peak optical depth is usually larger than the correction factor obtained using integrals of functions of the optical depth over velocity \citep{pine10}. The difference is especially substantial at high optical depth \citep{pine10}. However, the median peak optical depth of the C$^{18}$O (1-0) lines is 0.2 and the $^{13}$CO (1-0) lines without corresponding C$^{18}$O (1-0) emission is 0.5, indicating this problem with the opacity correction is not serious here.

The LTE assumption is another crucial factor of the uncertainties in determining the column density. In non-LTE case, the excitation temperature of the $^{13}$CO (1-0) and C$^{18}$O (1-0) lines may be very different from that of $^{12}$CO (1-0) \citep{liu12a}. We applied RADEX \citep{van07}, a one-dimensional non-LTE
radiative transfer code, to investigate the effect of non-LTE on the determination of column density of $^{13}$CO. The median value of $^{13}$CO column density under LTE assumption is 3.7$\times10^{15}$ cm$^{-2}$. The median values of the linewidth of $^{13}$CO (1-0) and $^{12}$CO (1-0) are 1.1 and 1.8 km~s$^{-1}$, respectively. We fix the volume density of H$_{2}$ as 2.0$\times10^{3}$ cm$^{-3}$, which is the mean value of the whole C3PO sample \citep{ade11}. Taking the typical values mentioned above and assuming that the relative abundance of $^{12}$CO to $^{13}$CO is 65, we simulated the emission of $^{12}$CO (1-0) and $^{13}$CO (1-0) in a parameter space for the kinetic temperature of [5,20] K using LVG model in RADEX.

In the left panel of Figure 1, we plot the excitation temperature of $^{12}$CO (1-0) and the ratio of the excitation temperature of $^{12}$CO (1-0) $T_{ex}^{12}$ to that of $^{13}$CO (1-0) $T_{ex}^{13}$ as a function of the kinetic temperature $T_{k}$. There is a linear relation between the the excitation temperature of $^{12}$CO (1-0) and the kinetic temperature: $T_{ex}^{12}=0.89\times T_{k}+0.73$. The ratio of $T_{ex}^{12}$ to $T_{ex}^{13}$ ranges from 1.13 to 1.26 with a mean value of 1.22, indicating that the LTE assumption overestimates the excitation temperature of $^{13}$CO (1-0) by a factor of $\sim$20\%. This leads to
an underestimation of the $^{13}$CO (1-0) opacity which in turn affects
the opacity correction of the column density. From the right panel of Figure 1, one can see that the optical depth of $^{12}$CO (1-0) decreases with the kinetic temperature but is much larger than 1. The optical depth of $^{13}$CO (1-0) calculated with RADEX also decreases with the kinetic temperature. At low kinetic temperature ($T_{k}<$8 K), the $^{13}$CO (1-0) emission may become optically thick. We also noticed that the optical depth of $^{13}$CO (1-0) calculated with RADEX is larger than that calculated under LTE assumption, especially at low kinetic temperatures. Therefore the opacity correction factor $C_{\tau}$ under non-LTE condition should be larger than $C_{\tau-LTE}$. However, the overestimation of the excitation temperature not only affect the optical depth but also affect the partition function. The partial function Z is given by
\begin{equation}
Z=\sum_{J=0}^{\infty}(2J+1)e^{\frac{-hB(J+1)}{kT_{ex}}}\approx\frac{kT_{ex}}{hB},~\textrm{when}~T_{ex}\gg hB/k.
\end{equation}
Thus the correction factor on Z due to non-LTE can be defined as $C_{Z}=\frac{Z_{non-LTE}}{Z_{LTE}}=\frac{T_{ex}^{13}}{T_{ex}^{12}}$. Where $T_{ex}^{13}$ and $T_{ex}^{12}$ are the excitation temperatures of $^{13}$CO (1-0) and $^{12}$CO (1-0), respectively. In the right panel of Figure 1, we plot $C_{Z}$ and $C_{\tau}C_{Z}/C_{\tau-LTE}$ as function of $T_{ex}^{12}$. We find that $C_{Z}$ is smaller than 1. $C_{\tau}C_{Z}/C_{\tau-LTE}$ is larger than 1 at lower temperature end, while smaller than 1 at high temperature end. However, $C_{\tau}C_{Z}/C_{\tau-LTE}$ varies slightly around 1 by a factor less than 20\%, indicating that the uncertainty  in the estimation of column density due to non-LTE effect is less than 20\%. For this reason, we use the column density estimated under LTE assumption in the following analysis.

The total column density of H$_{2}$ ($N_{H_{2}}$) can be calculated with \citep{ade11}
\begin{equation}
N_{H_{2}}=\frac{S_{\nu}}{\Omega_{c}\kappa_{\nu}B_{\nu}(T)\mu m_{H}}
\end{equation}
where S$_{\nu}$ is the flux density at 857 GHz integrated over the solid angle $\Omega_{c}=\frac{\pi}{4}\sigma_{Maj}\sigma_{Min}$ with $\sigma_{Maj}$ and $\sigma_{Min}$ the major and minor axis of the source, $B_{\nu}(T)$ is the Planck function at temperature T, $\mu=2.33$ is the mean molecular weight, and $m_{H}$ is the mass of atomic hydrogen. The dust opacity $\kappa_{\nu}$=0.1($\nu$/1 THz)$^{\beta}$~cm$^{2}$g$^{-1}$, with $\beta$ the dust emissivity spectral index \citep{ade11}.

The mass of the clumps are calculated as \citep{ade11}
\begin{equation}
M=\frac{S_{\nu}D^{2}}{\kappa_{\nu}B_{\nu}(T)}
\end{equation}
where S$_{\nu}$ is the integrated flux density at 857 GHz, D is the distance.

The Bolometric luminosity is defined by \citep{ade11}
\begin{equation}
L=4\pi D^{2}\int_{\nu}S_{\nu}d\nu
\end{equation}
where S$_{\nu}$ is the flux density at frequency $\nu$. The bolometric luminosity, L, is integrated over the frequency range 300 GHz$<\nu<$ 10 THz, using the modelled SEDs \citep{ade11}. The luminosity-to-mass ratio of the clumps ranges from $9\times10^{-3}$ to 3.5 L$_{\sun}$/M$_{\sun}$, with a median value of 0.2 L$_{\sun}$/M$_{\sun}$, indicating the Planck cold clumps are not significantly affected by star forming activities.

The averaged volume density is defined by
\begin{equation}
n_{H_{2}}=N_{H_{2}}/\sigma_{Min}
\end{equation}
The resulting volume density ranges from $\sim10^{2}$ to $\sim10^{5}$ cm$^{-3}$, with a mean value of 5.4$\times10^{3}$ cm$^{-3}$, which is slightly larger than the mean value (2$\times10^{3}$ cm$^{-3}$) of the whole C3PO sample \citep{ade11}.

If $^{12}$CO (1-0) emission has corresponding C$^{18}$O (1-0) emission, the column density of $^{12}$CO ($N_{^{12}CO}$) is converted from $N_{^{18}CO}$ with the $^{16}$O/$^{18}$O isotope ratio as \citep{wil94}
\begin{equation}
\frac{^{16}\textrm{O}}{^{18}\textrm{O}}=58.8\frac{\textrm{R}}{\textrm{kpc}}+37.1
\end{equation}
where R is the Galactocentric distance.

Otherwise, we converted $N_{^{13}CO}$ to $N_{^{12}CO}$ using the $^{12}C/^{13}C$ isotope ratio given by \citep{pine13}
\begin{equation}
\frac{^{12}\textrm{C}}{^{13}\textrm{C}}=4.7\frac{\textrm{R}}{\textrm{kpc}}+25.05
\end{equation}
The above relationship gives a $^{12}$C/$^{13}$C isotope ratio of 65 at R=8.5 kpc.

In the sample, about $\sim$30\% clumps have double or multiple velocity components in $^{12}$CO emission and $\sim$16\% have double or multiple velocity components in $^{13}$CO emission \citep{wu12}. We only considered the velocity components having $^{13}$CO emission in the calculation of the abundance. The observed $^{12}$CO abundance is [$^{12}$CO/H$_{2}$]=$\frac{\sum_{i}N_{^{12}CO}^{i}}{N_{T}}$, where i denotes the number of the $^{13}$CO velocity components of each source. One should keep in mind that in the calculation we assume the dust emission is uniform in the clumps, which may be not the case because in our CO mapping survey \citep{liu12b} and Herschel follow-up surveys \citep{juve10,juve12} most of the Planck clumps have sub-structures. This insufficiency can be improved in future by comparing the CO maps with the dust emission maps obtained from higher resolution observations (e.g. Herschel or ground-based telescopes like APEX).

The median and mean values of the observed gaseous $^{12}$CO abundance are 0.89$\times10^{-4}$ and 1.28$\times10^{-4}$, respectively.

\subsection{CO depletion and CO-to-H$_{2}$ conversion factor}

The CO depletion factor, $f_{D}$, is defined as:
\begin{equation}
f_{D}=\frac{X_{^{12}CO}^{E}}{[^{12}\textrm{CO}/\textrm{H}_{2}]}
\end{equation}
where $X_{^{12}CO}^{E}$ is the 'expected' abundance of CO.

Taking into account the variation of atomic carbon and oxygen abundances with the Galactocentric distance, the expected $^{12}$CO abundance at the Galactocentric distance R of each source is \citep{fon12}:
\begin{equation}
X_{^{12}CO}^{E}=8.5\times10^{-5}\textrm{exp}(1.105-0.13R(\textrm{kpc}))
\end{equation}
This relationship gives a canonical CO abundance of $\sim8.5\times10^{-5}$ in the neighborhood of the solar system \citep{fre82,lan89,pine08}.

The mean and median of CO depletion factor are 1.7 and 0.9, respectively. About 53\% Planck cold clumps have CO depletion factor smaller than 1. Only $\sim$13\% Planck cold clumps have CO depletion factor larger than 3 and Only about 5.6\% larger than 5. It seems that the CO gas in Planck cold clumps is not severely depleted. And due to the large beam size of PMO 13.7 m telescope and Planck satellite, we can not separate the depleted gas from the non-depleted gas, which should influence the interpretation of the CO depletion measurements. Thus our measurements indicate that on clump scale the CO depletion is not significant, in agreement with the fact that on such scales the observed emission arise mostly from low-density, non-depleted gas.

The CO-to-H$_{2}$ conversion factor $X_{CO-to-H_{2}}=\frac{N_{T}}{\sum_{i}I_{CO}^{i}}$, with $I_{CO}^{i}$ the integrated intensity of the $^{12}$CO (1-0) line. The median value of $X_{CO-to-H_{2}}$ for the whole sample is $2.8\times10^{20}$ cm$^{-2}$K$^{-1}$km$^{-1}$~s. However, CO emission may be saturated at high column density, which will add uncertainties in measuring $X_{CO-to-H_{2}}$. Within the Perseus molecular cloud, the $^{12}$CO emission saturates at $A_{V}\sim$4 mag \citep{pine08}. If we only consider the Planck cold clumps with $A_{V}<$4 mag (N$_{H_{2}}<3.8\times10^{21}$ cm$^{-2}$), the median and mean values of $X_{CO-to-H_{2}}$ are 1.7 and 1.9$\times10^{20}$ cm$^{-2}$K$^{-1}$km$^{-1}$~s, respectively, which are in agreement with the mean value of (1.8$\pm0.3)\times10^{20}$ cm$^{-2}$K$^{-1}$km$^{-1}$~s for the Milky Way \citep{dame01}.

\section{Discussion}
\subsection{The relationships between various physical parameters}
In Figure 2, we investigate the relationships between CO abundance and dust temperature (T$_{d}$), dust emissivity spectral index ($\beta$), column density of H$_{2}$ (N$_{H_{2}}$), volume density of H$_{2}$ (n$_{H_{2}}$), luminosity-to-mass ratio (L/M) as well as the non-therm one dimensional velocity dispersion ($\sigma_{NT}$). CO abundance strongly correlates with T$_{d}$ and L/M and anti-correlates with $\beta$, N$_{H_{2}}$ and n$_{H_{2}}$. These relationships were well fitted with a power-law function ($y=\eta\cdot x^{\zeta}$). The coefficients of each model as well as the correlation coefficients $R^{2}$ are displayed in the upper-right corner of each panel. There is no correlation between CO abundance and $\sigma_{NT}$, indicating that turbulence has no effect on the fluctuation of CO abundance. The lower CO abundance for the clumps with lower T$_{d}$ and with higher N$_{H_{2}}$ or n$_{H_{2}}$ indicates that CO gas may freeze out significantly towards cold and dense regions. The growth of icy mantles on dust grains could steepen the slope of the dust SED and thus increase the emissivity spectral index $\beta$ \citep{sch10}. The anti-correlation between CO abundance and $\beta$ indicates that huge amount of gaseous CO transforms to icy CO with the growth of icy mantles on dust grains. However, the presence of observational errors makes the anti-correlation between the T$_{d}$ and $\beta$ become unreliable in the C3PO sample \citep{ade11}. Thus the relationships between $\beta$ and the other physical parameters should be taken seriously and tested by further more detailed observations.

In Figure 3, we plot the CO depletion factor f$_{D}$ as a function of T$_{d}$, $\beta$,  N$_{H_{2}}$, n$_{H_{2}}$, L/M and $\sigma_{NT}$. The f$_{D}$ significantly anti-correlates with T$_{d}$ and L/M and positively correlates with $\beta$, N$_{H_{2}}$ and n$_{H_{2}}$. These relationships were well fitted with a power-law function ($y=\eta\cdot x^{\zeta}$). In each panel, we divide the data into ten bins and plot median f$_{D}$ in each bin as red filled circles. We find that the median f$_{D}$ is larger than 1 in bins with T$_{d}<$10.8 K or $\beta>2.5$ or N$_{H_{2}}>$6.6$\times10^{21}$ cm$^{-2}$ or n$_{H_{2}}>$1.7$\times10^{3}$ cm$^{-3}$ or L/M$<$0.16, indicating that CO gas freeze out in cold and dense regions without significant internal heating. There is also no correlation between CO depletion factor f$_{D}$ and $\sigma_{NT}$.

The variation of gaseous CO abundance in molecular clouds should seriously hamper its utility as an estimator of the total hydrogen column density. In Figure 4, we plot the CO-to-H$_{2}$ conversion factor $X_{CO-to-H_{2}}$ against T$_{d}$, $\beta$,  N$_{H_{2}}$, n$_{H_{2}}$, L/M and $\sigma_{NT}$. In each panel, the median values of $X_{CO-to-H_{2}}$ in each bin are plotted as red filled circles. There is an inverse correlation between $X_{CO-to-H_{2}}$ and T$_{d}$ and L/M. While $X_{CO-to-H_{2}}$ positively correlates to $\beta$, N$_{H_{2}}$ and n$_{H_{2}}$. These relationships can be well fitted with power-law functions. There is no correlation between $X_{CO-to-H_{2}}$ and $\sigma_{NT}$. We found that the median value of $X_{CO-to-H_{2}}$ is larger than 2$\times10^{20}$ cm$^{-2}$K$^{-1}$km$^{-1}$~s in bins with T$_{d}<$12 K or $\beta>2.4$ or N$_{H_{2}}>$2.6$\times10^{21}$ cm$^{-2}$ or n$_{H_{2}}>$2.8$\times10^{2}$ cm$^{-3}$ or L/M$<$0.34.

\subsection{Gaseous CO abundance --- An evolutionary tracer for molecular clouds}
The freezing out of gaseous CO onto grain surfaces strongly influences the physical and chemical properties of molecular clouds. As a major destroyer of molecular ions, the CO depletion leads to a change in the relative abundances of major charge carriers (e.g.H$_{3}^{+}$, N$_{2}$H$^{+}$ and HCO$^{+}$) and thus causes variation in the ionizing degree \citep{case99,berg07}. Another second-effect induced by CO depletion is deuterium enrichments in cold cores \citep{case99,berg07}. These effects make gaseous CO abundance a promising tool to time the evolution of molecular clouds. As discussed in section 1, significant fraction of CO molecules is transformed from gas to solid as the gas density increases \citep{her11,whi10}. As molecular clump evolves, the density and temperature increase. The bolometric luminosity also increases as the protostars form and evolve in molecular clumps \citep{Emp09}. The ratio of bolometric luminosity to submillimeter emission is also used as an effective indicator for protostar evolution. In this work, we found that the relative abundance of gaseous CO
significantly anti-correlates with dust temperature and luminosity-to-mass ratio and positively correlates with column density, volume density and dust emissivity spectral index, indicating that gaseous CO abundance can also well serve as an evolutionary indicator.

One should keep in mind that the Planck cold clumps are cold ($<14$ K) \citep{ade11}, turbulence dominated and have relatively low column densities comparing with the other star forming regions \citep{wu12}. They are mostly quiescent and lacking star forming activities, indicating that the Planck cold clumps are most likely at the very initial evolutionary stages of molecular clouds \citep{wu12}. Thus the relationships between various physical parameters reported here may be only valid for molecular clouds without significant star forming activities. However, our results indicate that gaseous CO abundance (or depletion) can be used as a tracer for the evolution of molecular clouds. Actually, people have already used CO depletion to distinguish relatively evolved starless cores from more-recently condensed cores \citep{taf04,taf10}.

\section*{Acknowledgment}
\begin{acknowledgements}
We are grateful to the staff in the Qinghai Station of Purple Mountain Observatory. Thanks for the Key Laboratory for Radio Astronomy, CAS for partly
support the telescope operating. This work was funded by China Ministry of Science and Technology under State Key Development Program for Basic Research 2012CB821800 and by the National Natural Science Foundation of China under
Grant No.11073003 and 11233004. Many thanks to the anonymous
referee for his/her useful suggestions and comments.
\end{acknowledgements}

\begin{figure}
\begin{minipage}[c]{0.5\textwidth}
  \centering
  \includegraphics[width=80mm,height=65mm,angle=0]{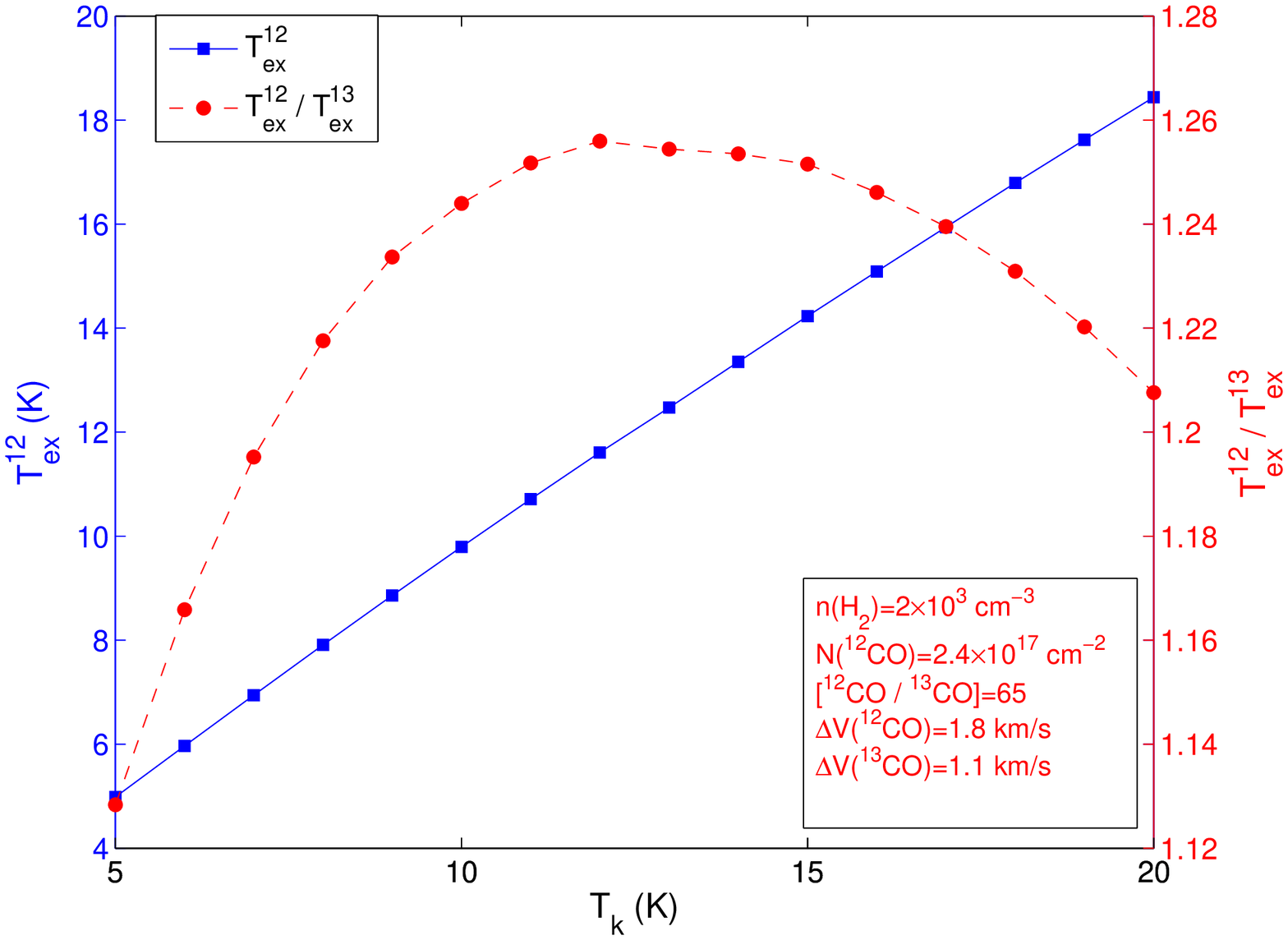}
\end{minipage}
\begin{minipage}[c]{0.5\textwidth}
  \centering
  \includegraphics[width=80mm,height=65mm,angle=0]{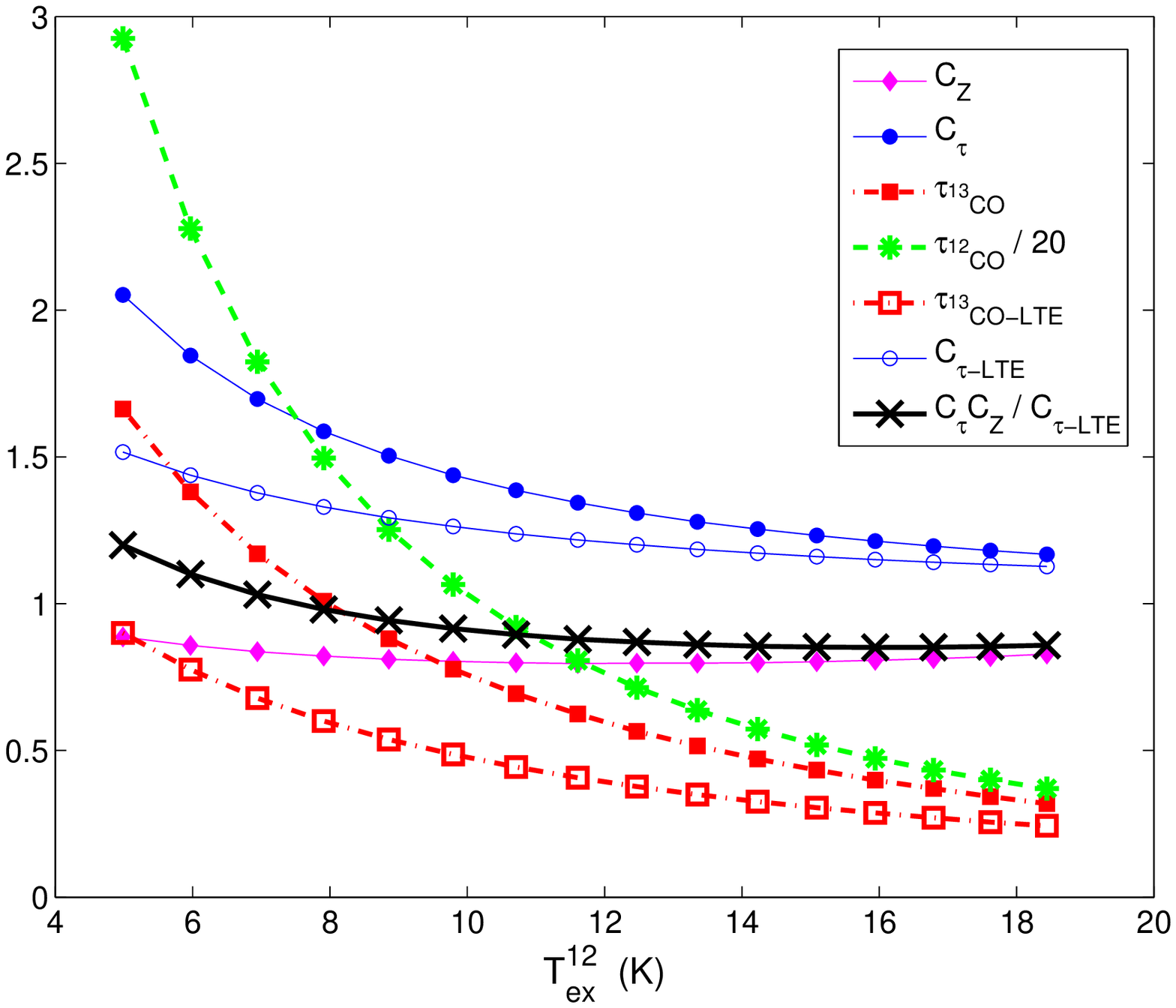}
\end{minipage}
\caption{Left: $T_{ex}^{12}$ and $T_{ex}^{12}/T_{ex}^{13}$ as a function of $T_{k}$ in the simulation with RADEX. Right: Optical depth and correction factors (see section 3.1) from RADEX simulation. }
\end{figure}

\begin{figure}
\begin{minipage}[c]{0.5\textwidth}
  \centering
  \includegraphics[width=80mm,height=65mm,angle=0]{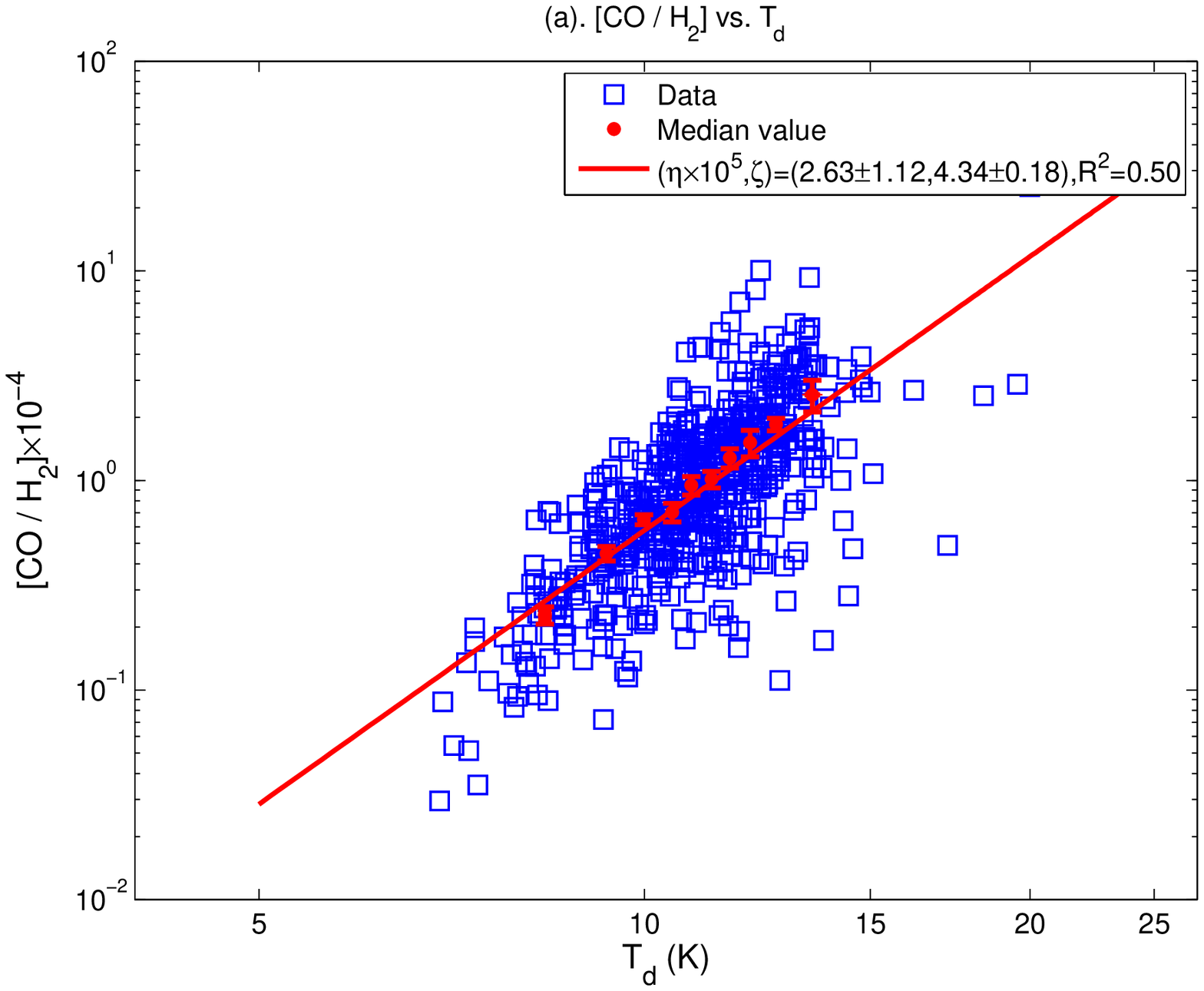}
\end{minipage}
\begin{minipage}[c]{0.5\textwidth}
  \centering
  \includegraphics[width=80mm,height=65mm,angle=0]{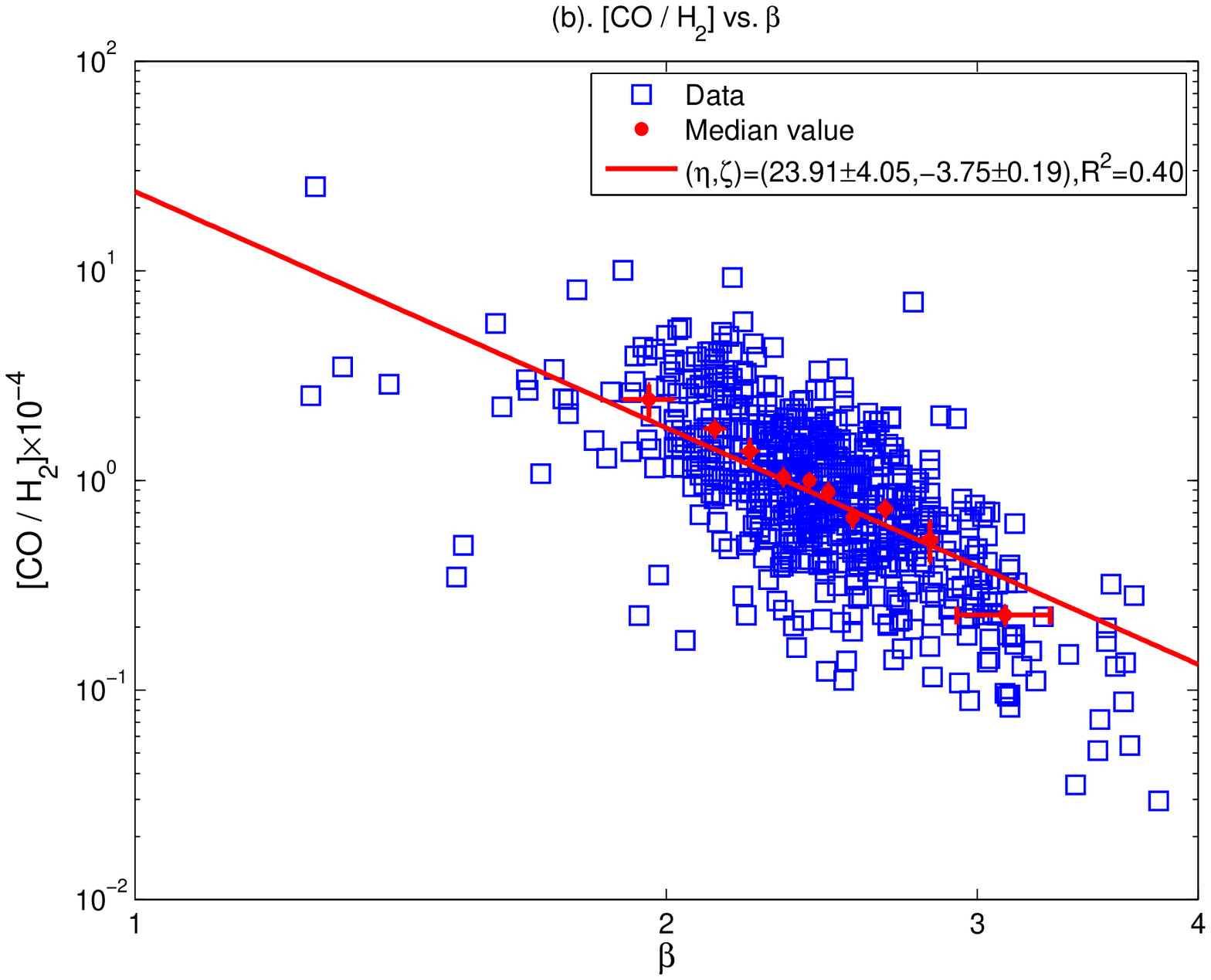}
\end{minipage}
\begin{minipage}[c]{0.5\textwidth}
  \centering
  \includegraphics[width=80mm,height=65mm,angle=0]{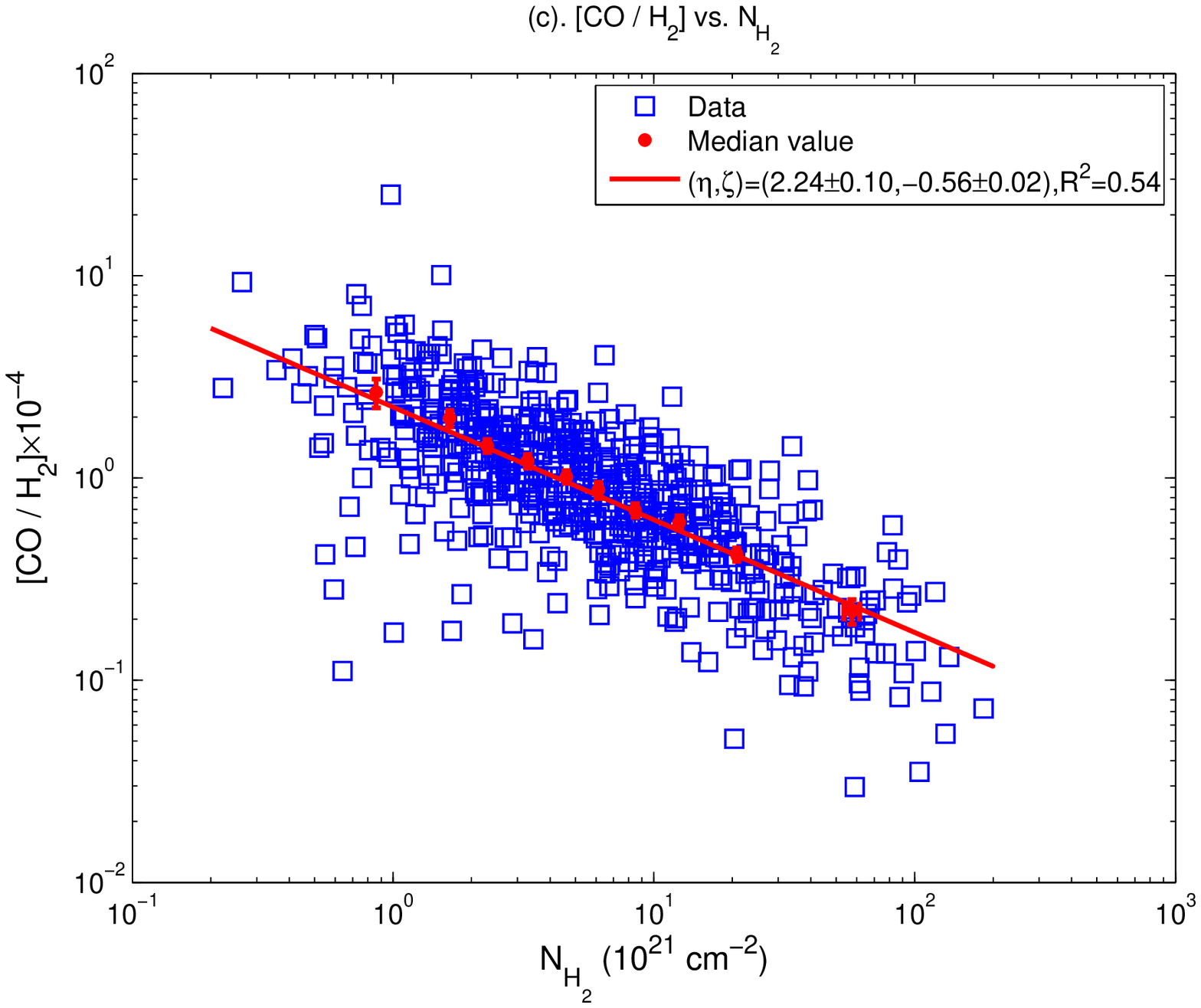}
\end{minipage}
\begin{minipage}[c]{0.5\textwidth}
  \centering
  \includegraphics[width=80mm,height=65mm,angle=0]{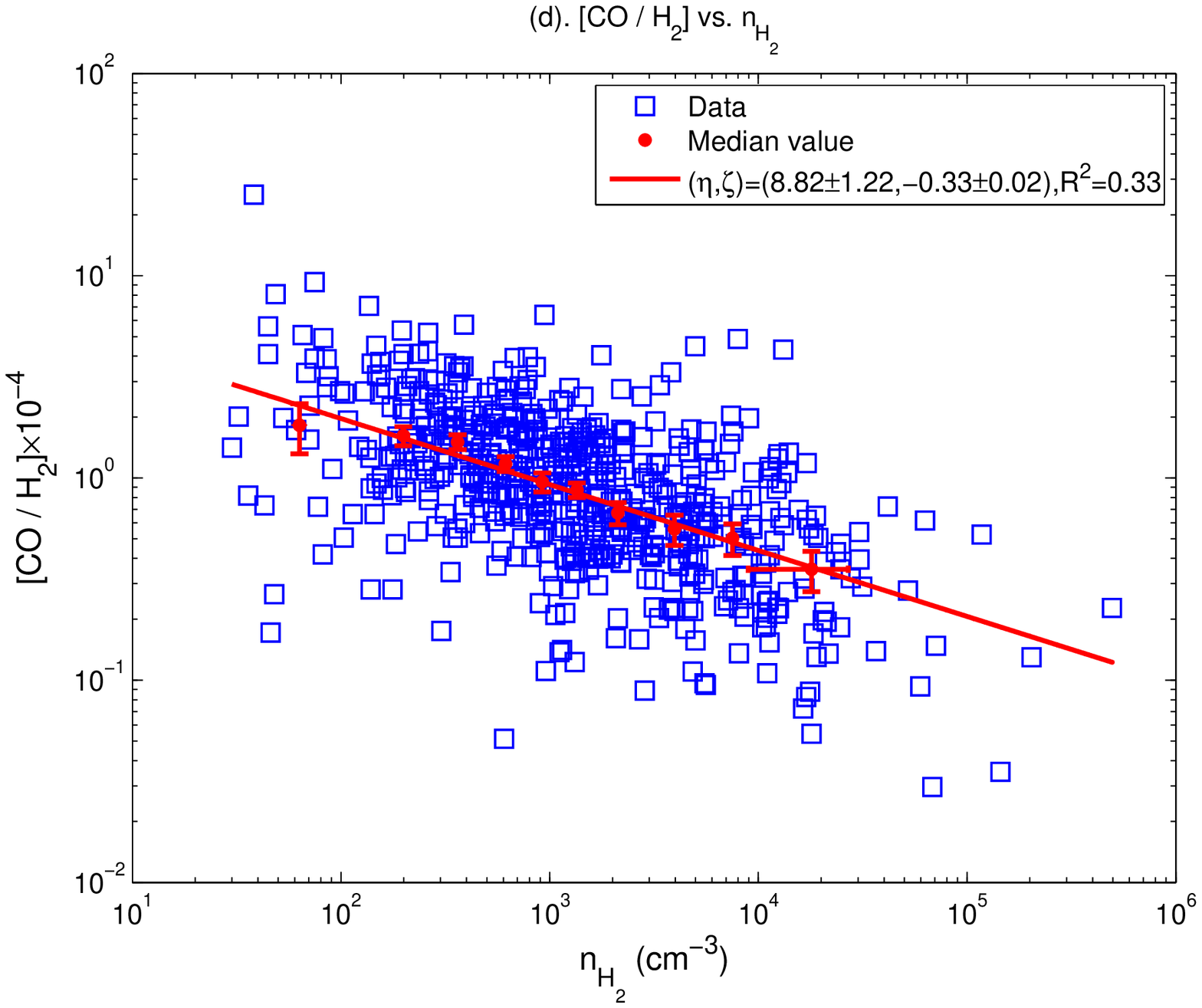}
\end{minipage}
\begin{minipage}[c]{0.5\textwidth}
  \centering
  \includegraphics[width=80mm,height=65mm,angle=0]{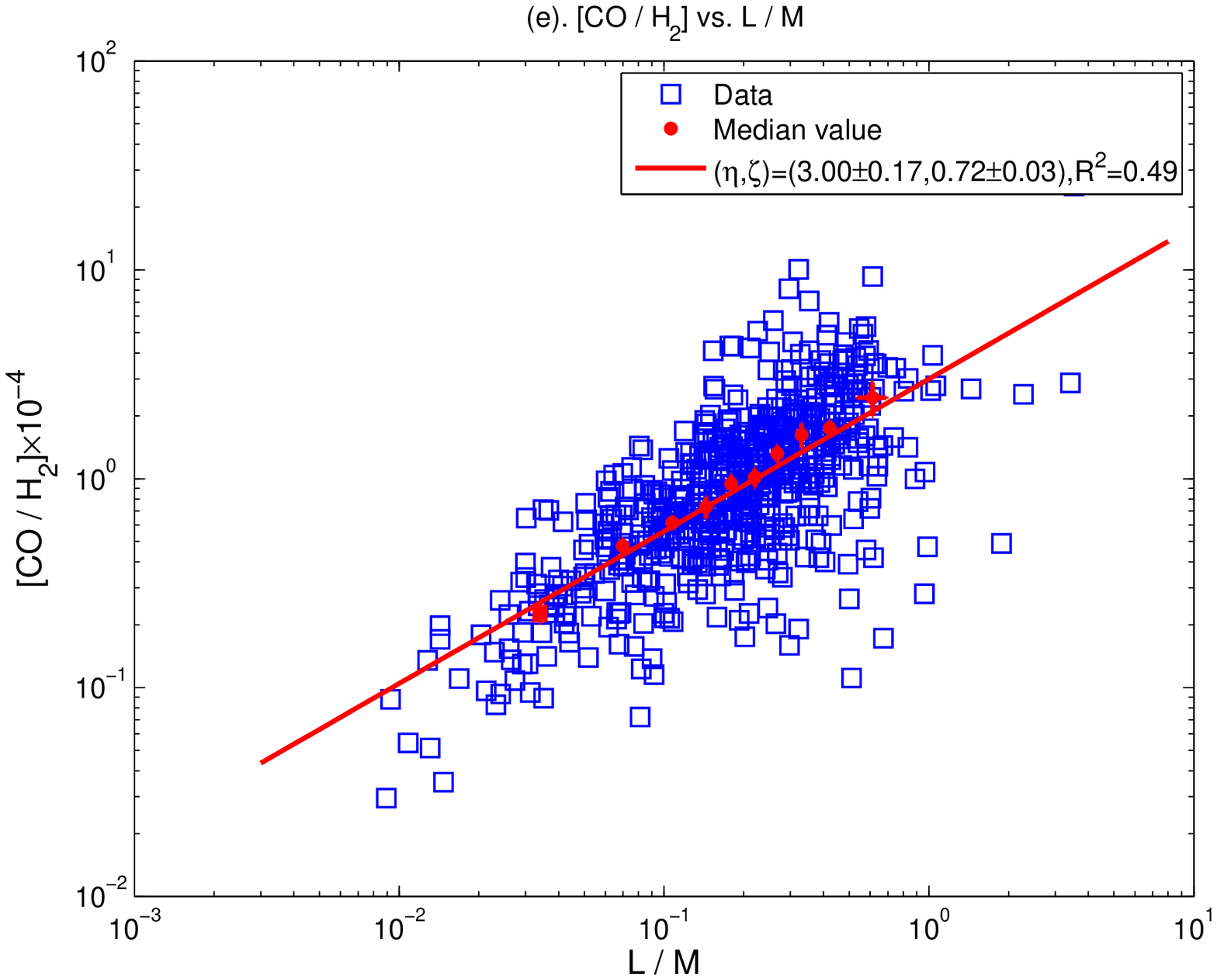}
\end{minipage}
\begin{minipage}[c]{0.5\textwidth}
  \centering
  \includegraphics[width=80mm,height=65mm,angle=0]{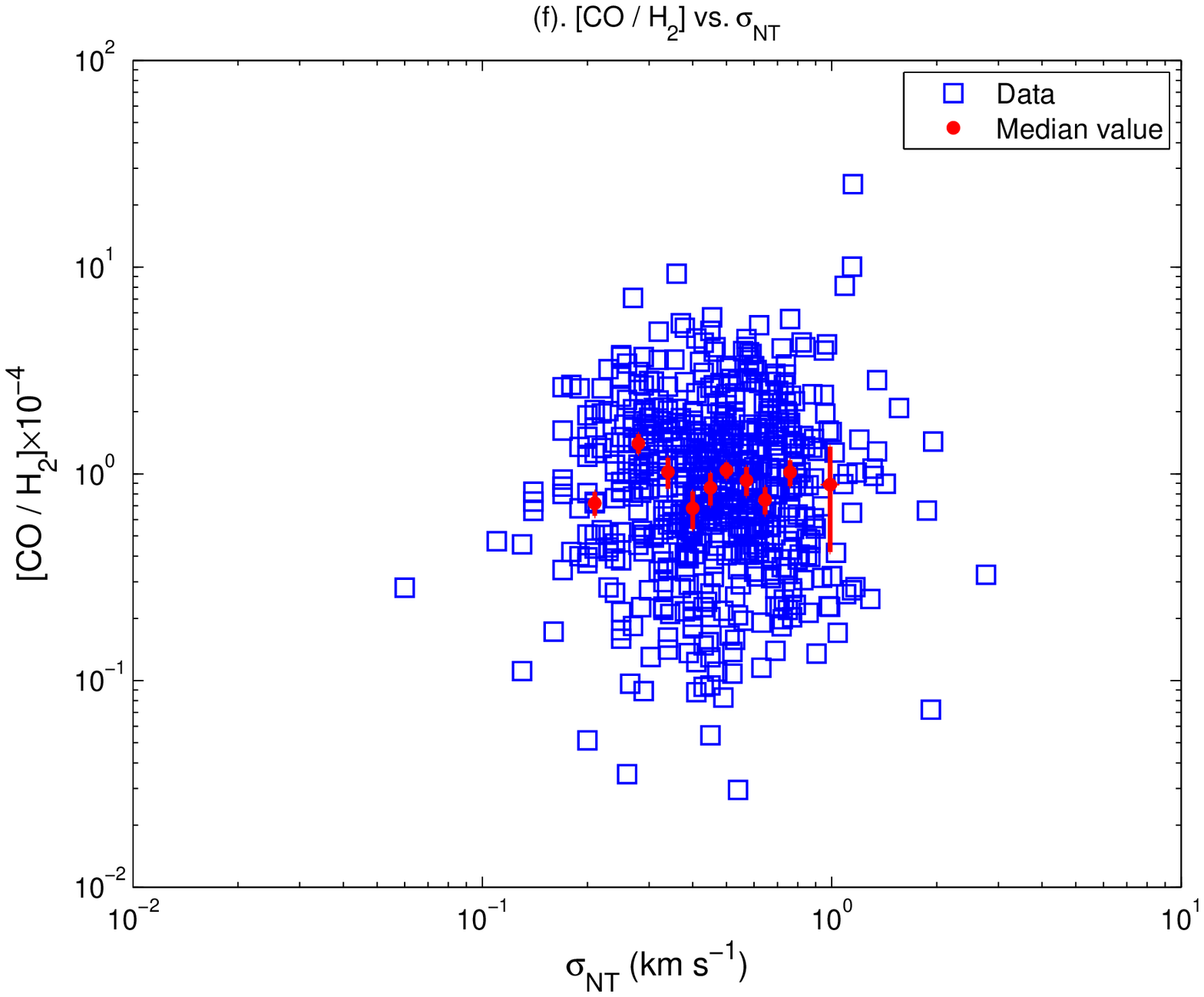}
\end{minipage}
\caption{The CO relative abundance [CO/H$_{2}$] as a function of T$_{d}$ (a), $\beta$ (b), N$_{H_{2}}$ (c), n$_{H_{2}}$ (d), L/M (e) and $\sigma_{NT}$ (f). The red line in each panel (also in Figure 3 and 4) is power-law fitting. In each panel (also in Figure 3 and 4), the red filled circles represent the median value in each bin and the size of the green error bars represent the standard error.  }
\end{figure}

\begin{figure}
\begin{minipage}[c]{0.5\textwidth}
  \centering
  \includegraphics[width=80mm,height=65mm,angle=0]{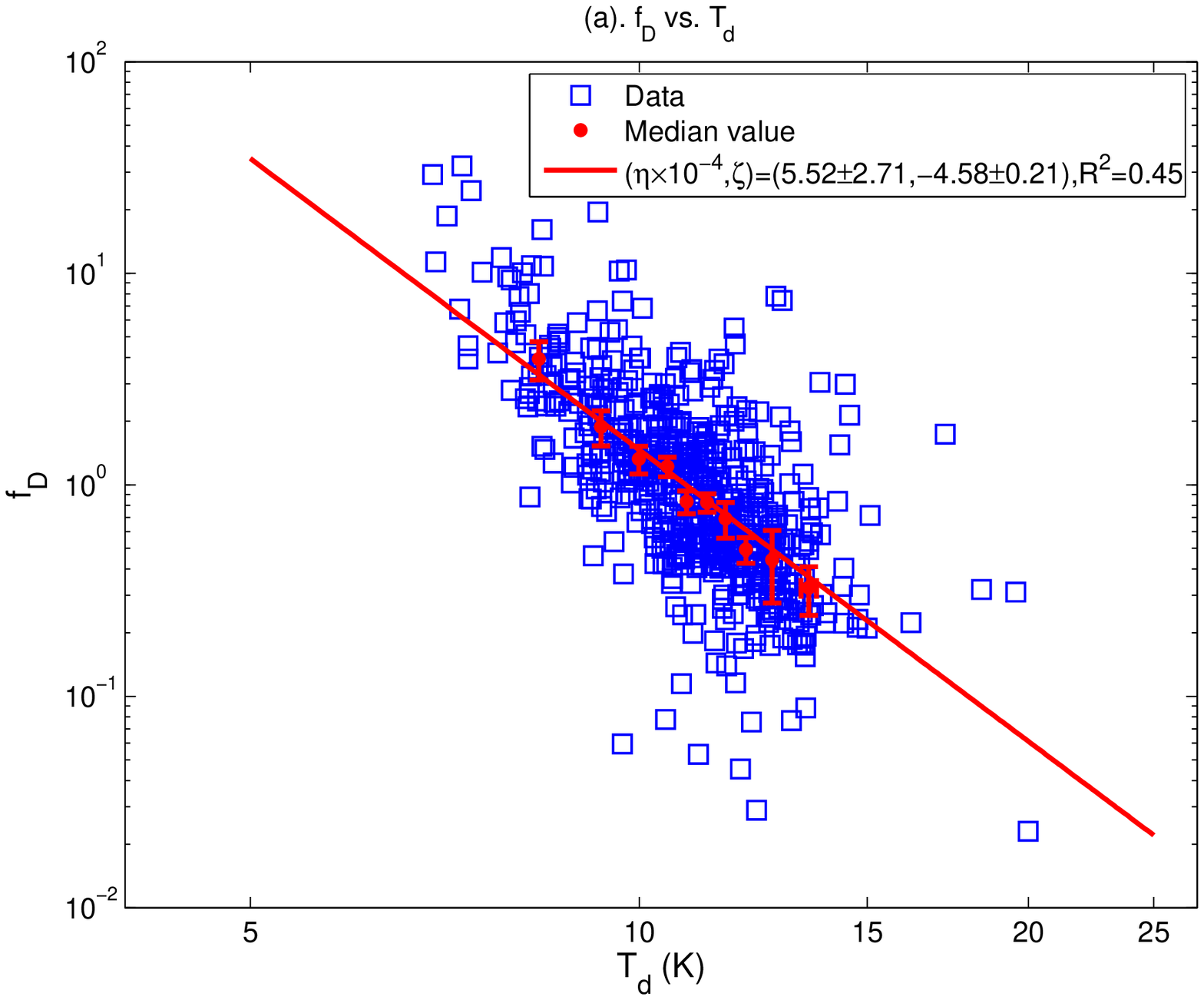}
\end{minipage}
\begin{minipage}[c]{0.5\textwidth}
  \centering
  \includegraphics[width=80mm,height=65mm,angle=0]{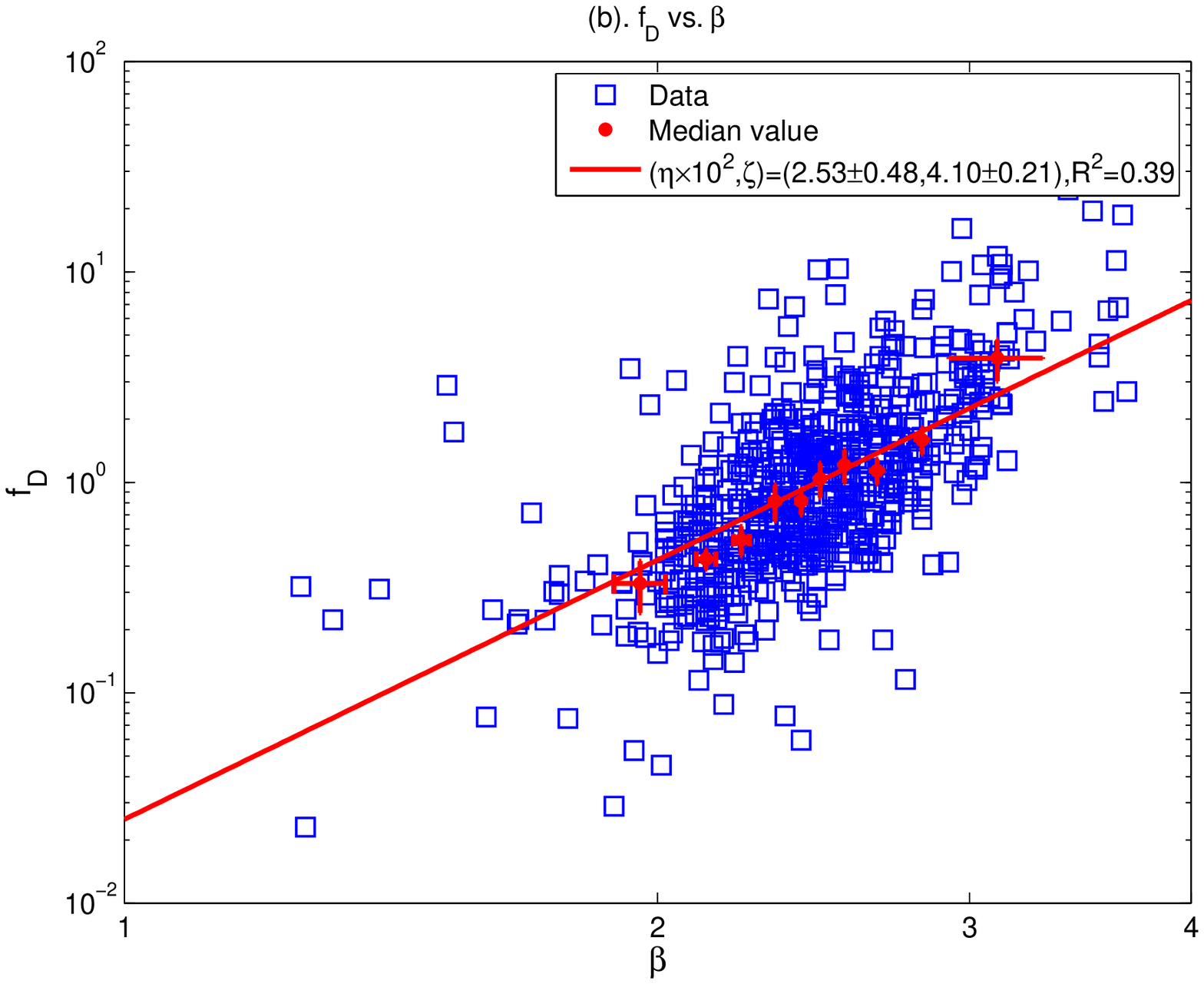}
\end{minipage}
\begin{minipage}[c]{0.5\textwidth}
  \centering
  \includegraphics[width=80mm,height=65mm,angle=0]{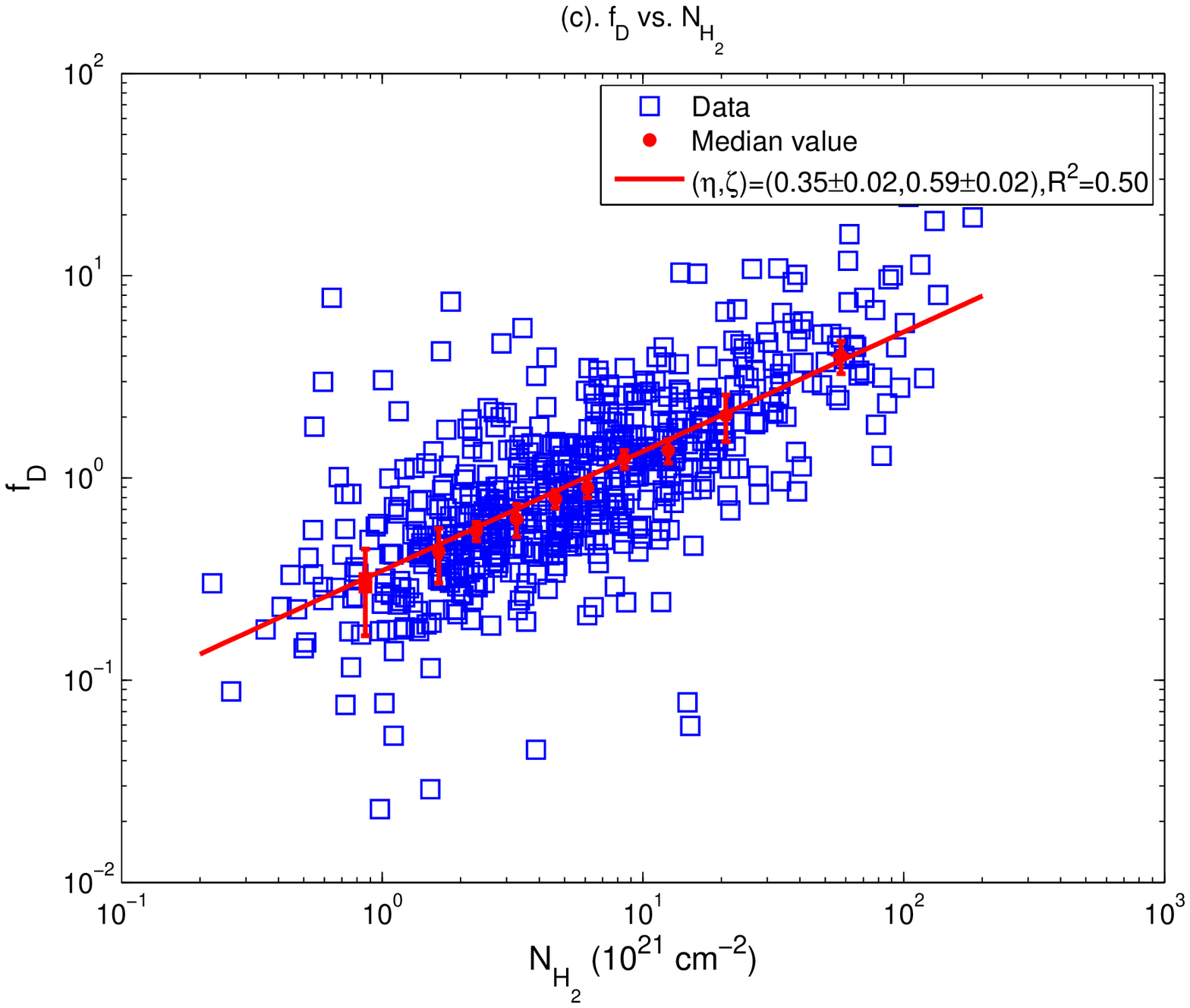}
\end{minipage}
\begin{minipage}[c]{0.5\textwidth}
  \centering
  \includegraphics[width=80mm,height=65mm,angle=0]{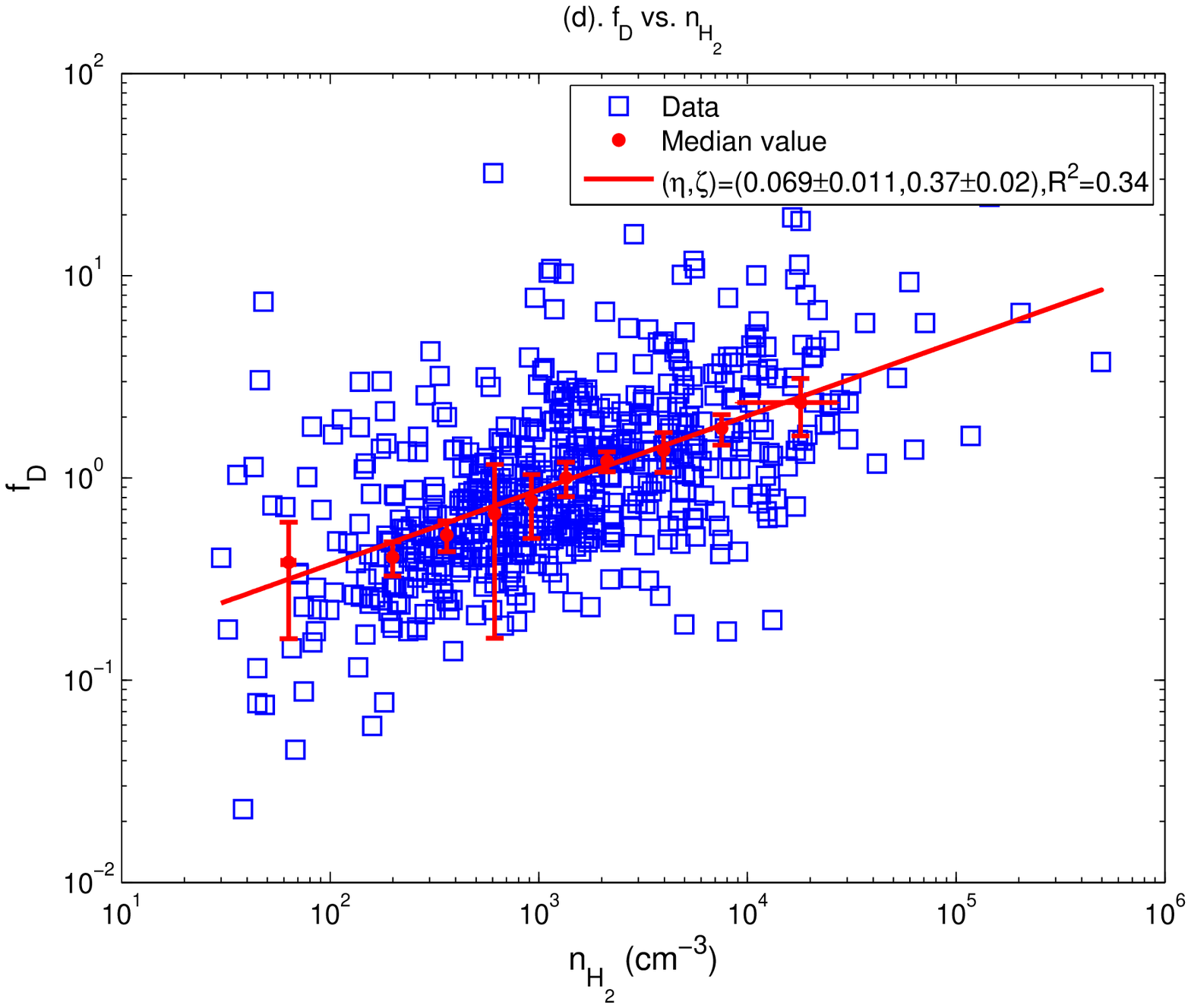}
\end{minipage}
\begin{minipage}[c]{0.5\textwidth}
  \centering
  \includegraphics[width=80mm,height=65mm,angle=0]{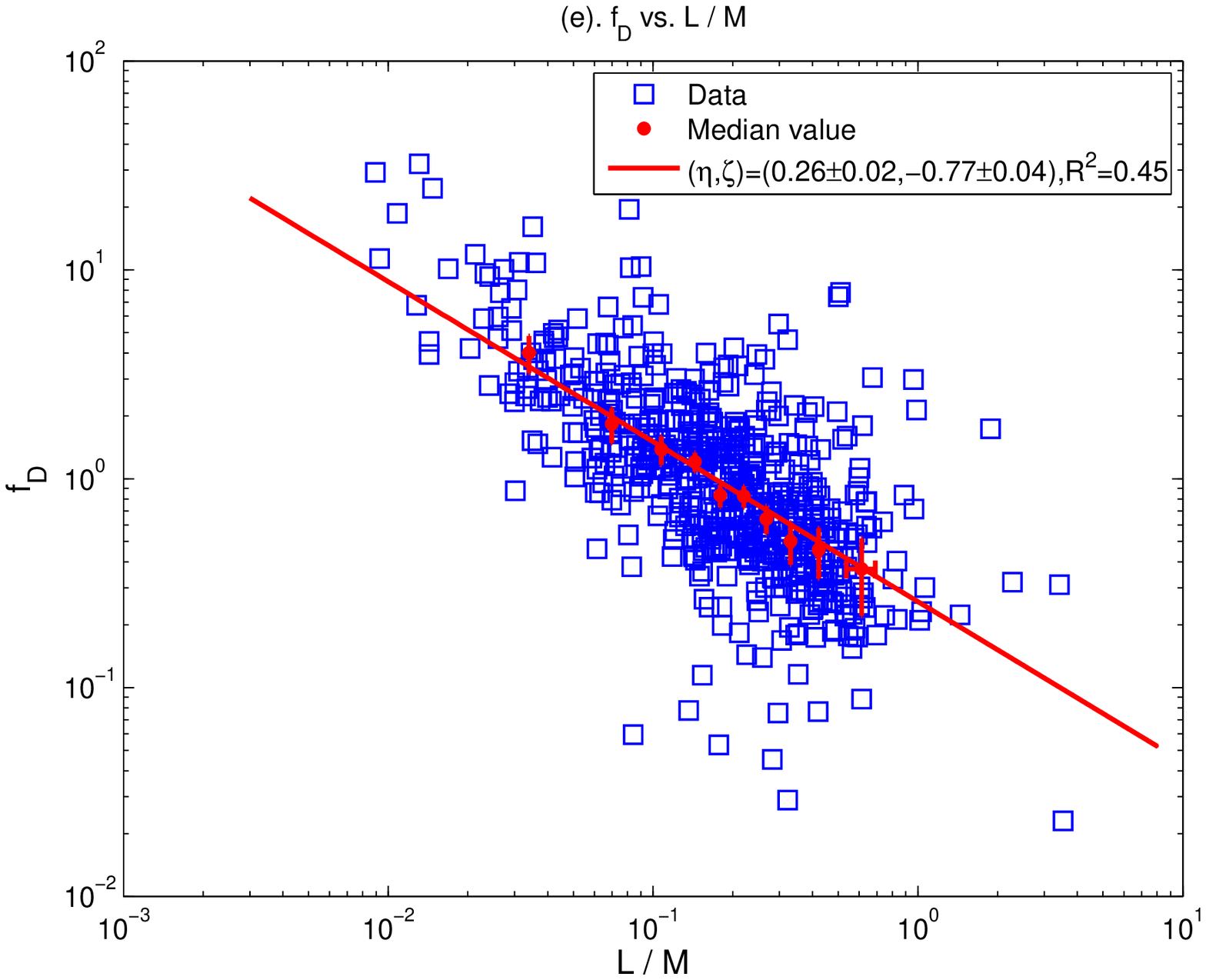}
\end{minipage}
\begin{minipage}[c]{0.5\textwidth}
  \centering
  \includegraphics[width=80mm,height=65mm,angle=0]{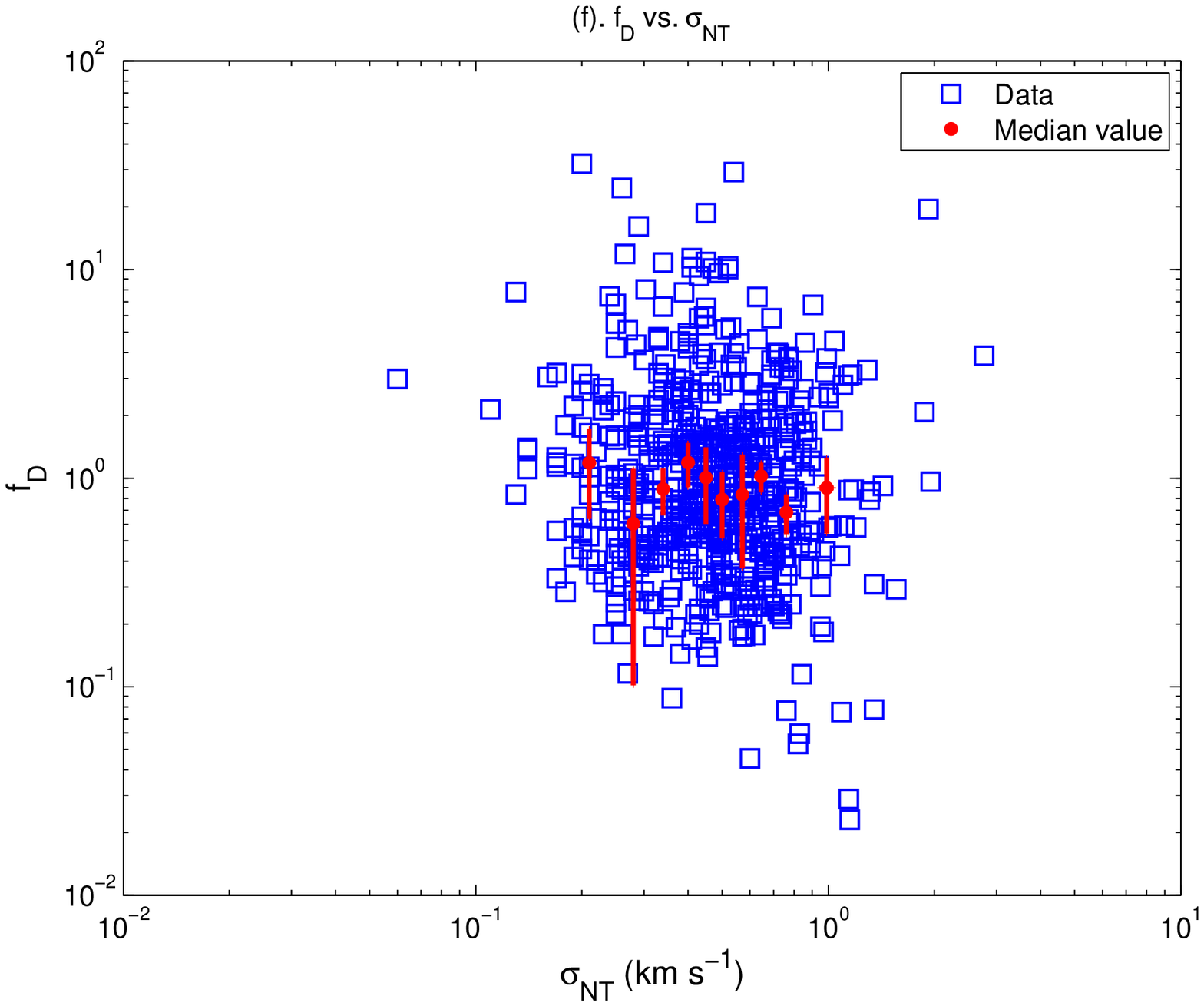}
\end{minipage}
\caption{The CO depletion factor f$_{D}$ as a function of T$_{d}$ (a), $\beta$ (b), N$_{H_{2}}$ (c), n$_{H_{2}}$ (d), L/M (e) and $\sigma_{NT}$ (f). Other signs are the same as in Figure 2.}
\end{figure}

\begin{figure}
\begin{minipage}[c]{0.5\textwidth}
  \centering
  \includegraphics[width=80mm,height=65mm,angle=0]{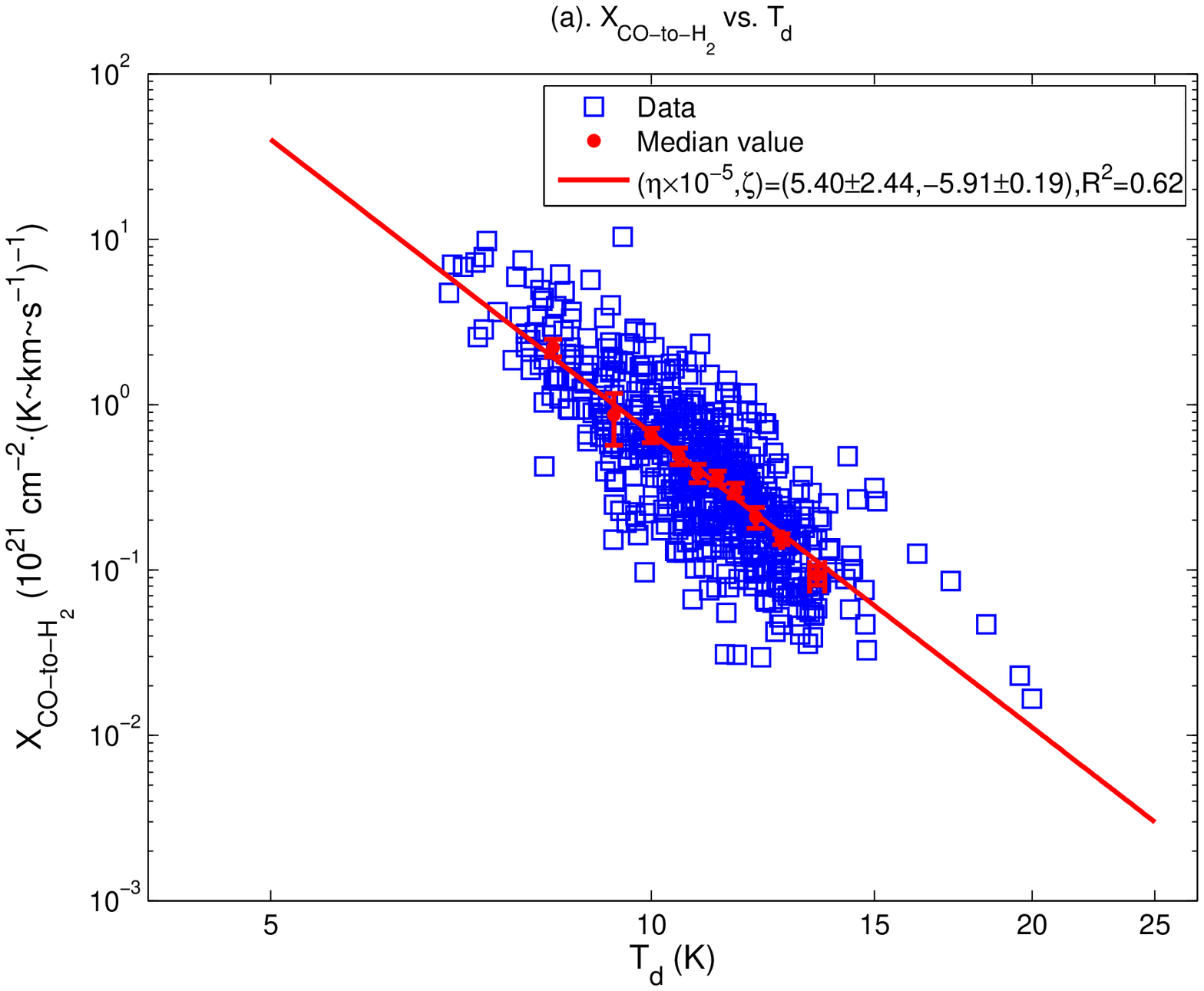}
\end{minipage}
\begin{minipage}[c]{0.5\textwidth}
  \centering
  \includegraphics[width=80mm,height=65mm,angle=0]{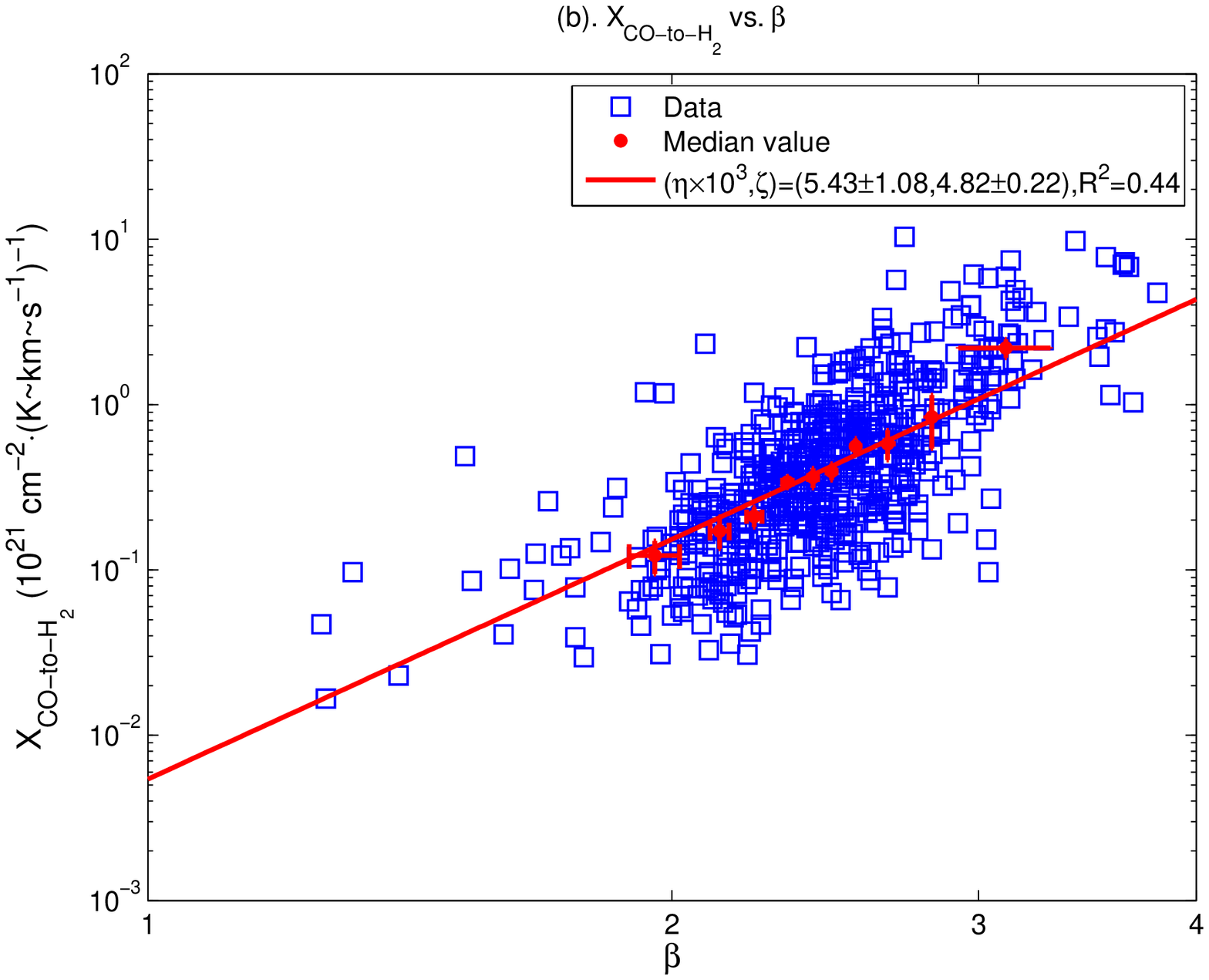}
\end{minipage}
\begin{minipage}[c]{0.5\textwidth}
  \centering
  \includegraphics[width=80mm,height=65mm,angle=0]{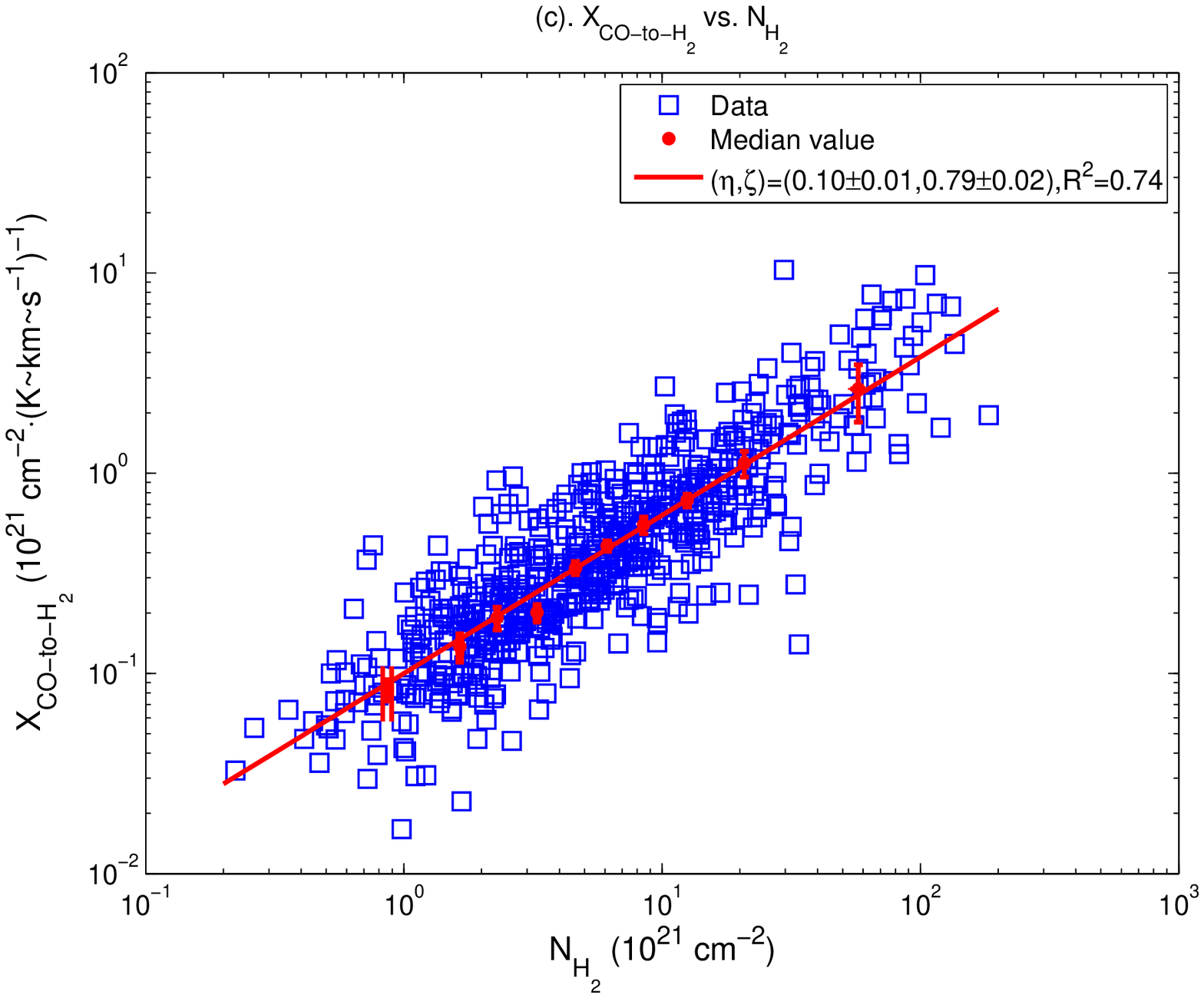}
\end{minipage}
\begin{minipage}[c]{0.5\textwidth}
  \centering
  \includegraphics[width=80mm,height=65mm,angle=0]{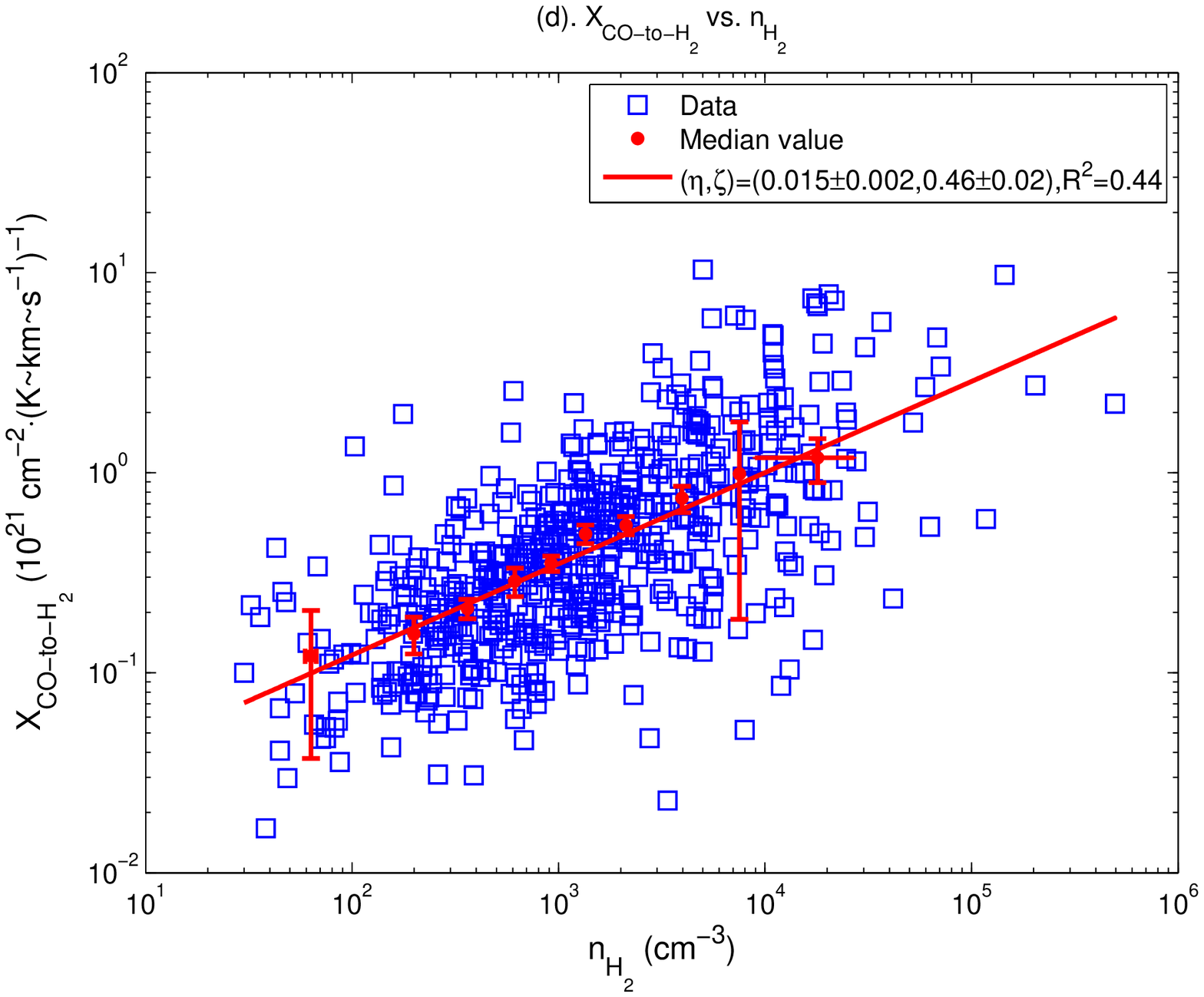}
\end{minipage}
\begin{minipage}[c]{0.5\textwidth}
  \centering
  \includegraphics[width=80mm,height=65mm,angle=0]{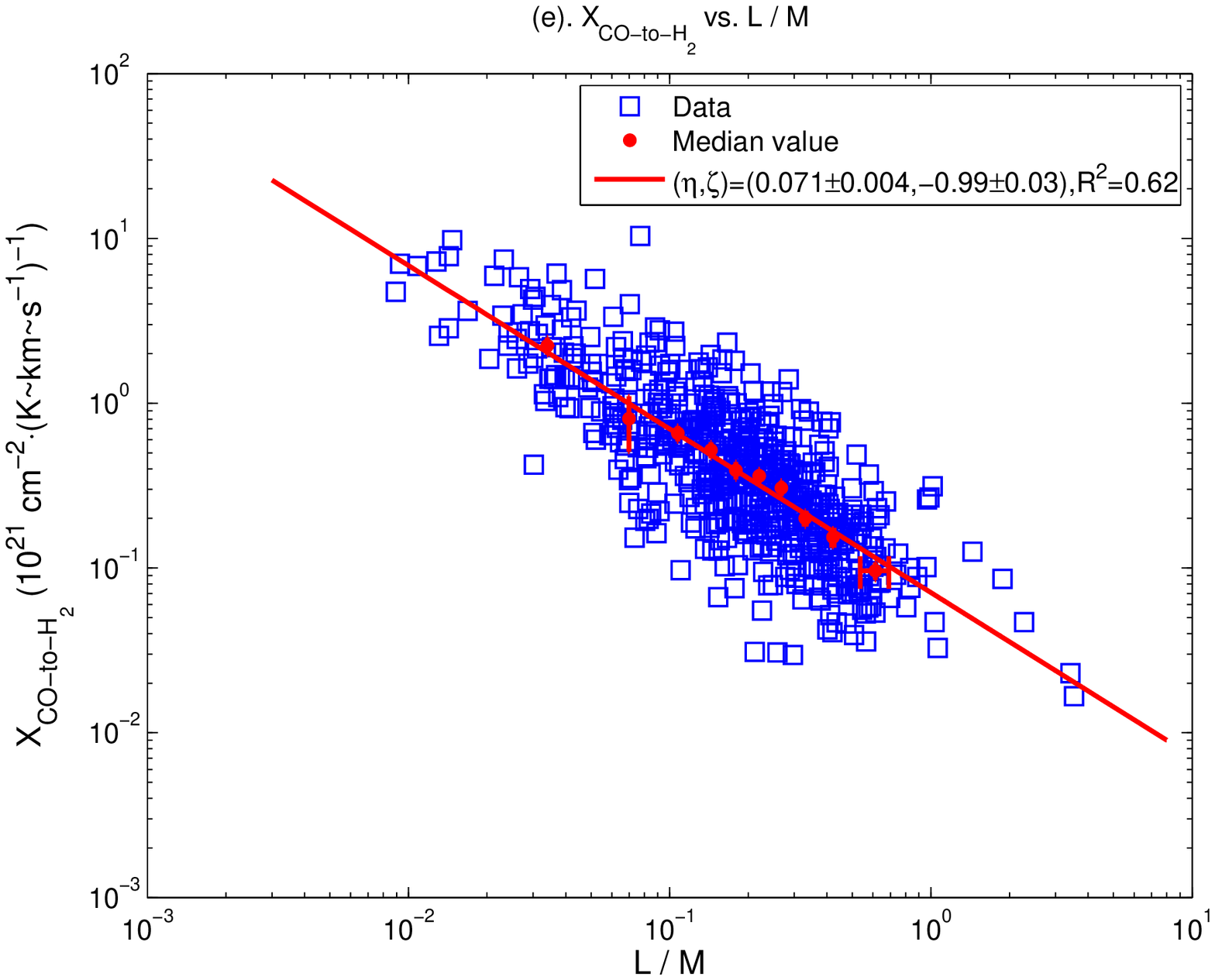}
\end{minipage}
\begin{minipage}[c]{0.5\textwidth}
  \centering
  \includegraphics[width=80mm,height=65mm,angle=0]{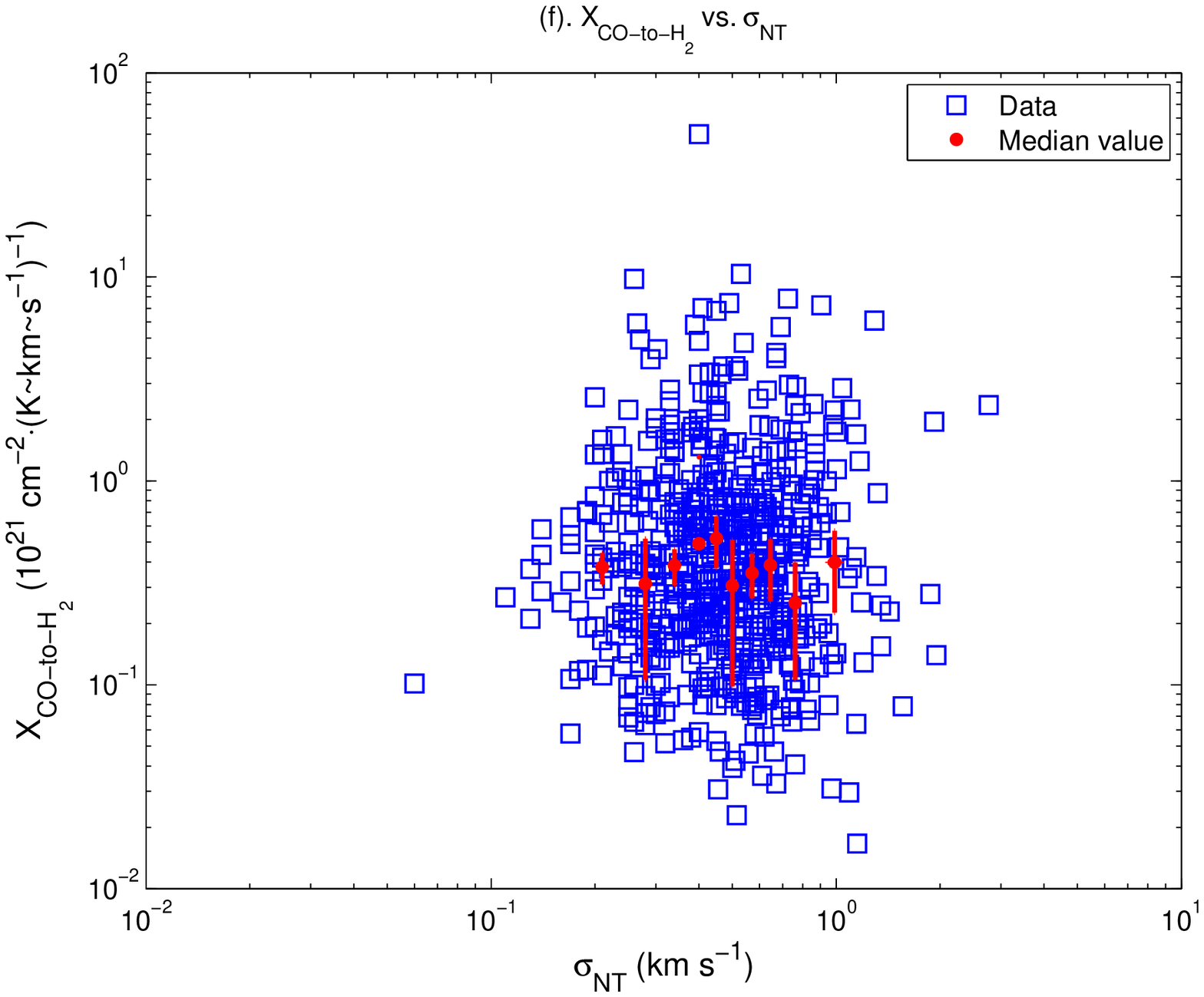}
\end{minipage}
\caption{The CO-to-H$_{2}$ conversion factor X$_{CO-to-H_{2}}$ as a function of T$_{d}$ (a), $\beta$ (b), N$_{H_{2}}$ (c), n$_{H_{2}}$ (d), L/M (e) and $\sigma_{NT}$ (f). Other signs are the same as in Figure 2.}
\end{figure}

\end{document}